\documentclass[preprint]{rsl}
\title{Developing a Real Time Processing System for HERA}
\author{Paul La Plante, Peter K. G. Williams, Joshua S. Dillon}

\begin{document}

\maketitle

%
% Manuscript contents begins.
%

\begin{abstract}
  The Hydrogen Epoch of Reionization Array (HERA) is a radio telescope in the
  Karoo desert of South Africa endeavoring to observe Cosmic Dawn and the Epoch
  of Reionization. When fully constructed, it will consist of 350 antennas and
  generate over 60 terabytes (TB) of data each night. In order to keep pace with
  the relatively large rate of data, we have developed the real-time processing
  (RTP) system for HERA. The RTP is responsible for inspecting data for
  acceptable levels of quality and flagging unusable data, as well as providing
  an initial calibration solution. The RTP system consists of the pipeline
  management system, called the \texttt{hera\_opm} package, and the data
  management system, called the \texttt{librarian}. Though the systems have been
  developed in the context of analyzing HERA data, they feature public and open
  source software, and have been designed to be adapted and used in other
  contexts as necessary.
\end{abstract}

\section{Introduction}
As current- and next-generation telescopes are constructed, they are projected
to generate more data than ever before. This fact is particularly true for the
field of radio astronomy, where the volume of interferometric datasets are among
the largest in astronomy. Part of the reason for the large data comes from the
availability of modern digital signal processing technology, where the number of
array elements and spectral resolution can quickly lead to total data volumes
orders of magnitude larger than, e.g., optical astronomy. The raw datasets are
known as visibilities, and encode the degree of correlation observed by a pair
of antennas at a given time of observation in the form of a complex-valued
quantity. For a given array of $N_\mathrm{ant}$ antennas, one can construct
$N_\mathrm{ant}(N_\mathrm{ant}+1)/2$ unique pairs of antennas (including
auto-correlations). Each of these baselines produces a spectrum containing
$N_\mathrm{freq}$ frequency channels every $t_\mathrm{corr}$ seconds. A spectrum
is produced for each instrumental polarization element, and combined to generate
$N_\mathrm{pol}$ spectra. For a given observation of length $T_\mathrm{obs}$,
the number of snapshots generated in time is
$N_\mathrm{times} = T_\mathrm{obs}/t_\mathrm{corr}$, so the total number of
spectra is
$N_\mathrm{times}N_\mathrm{pol}N_\mathrm{ant}(N_\mathrm{ant}+1)/2$. Visibilities
are typically stored as single-precision floating point numbers for both the
real and imaginary components. Thus, the total volume of data produced $V$ is:
\begin{equation}
V = 4 N_\mathrm{ant} (N_\mathrm{ant} + 1)N_\mathrm{freq} N_\mathrm{pol} N_\mathrm{times}\, \mathrm{bytes}.
\end{equation}
When fully constructed, HERA\footnote{https://reionization.org}
\cite{deboer_etal2017} will consist of 350 antennas, produce spectra with 6144
frequency channels, generate 4 instrumental polarizations, and have a
correlation sampling rate of 2 seconds. Thus, for a typical observation of 12
hours, this leads to a raw data volume of nearly 240 terabytes (TB). The actual
data volume produced by HERA is expected to be smaller than this by several
factors, primarily by effectively averaging data longer for shorter baselines
using techniques proposed in the literature
\cite{wijnholds_etal2018}. Nevertheless, the nightly volume is projected to be
over 60 TB. Obviously, analyzing, calibrating, and cataloging such a large
volume of data manually each night would be difficult, and storing the raw data
indefinitely would requite a tremendous amount of storage.

To accommodate the large volume of data, the HERA collaboration has developed
the real-time processing (RTP) system, designed to keep pace with data
acquisition rates and reduce the data to a reasonable volume in a short amount
of time. The goal is to provide data quality metrics, calibration solutions, and
visualizations of data in near-real time, and subsequently reduce the volume of
data to a quantity that is acceptable for long-term storage. Once the data have
been stored on site, we also want to transfer the files to remote processing
clusters, which can make the data widely available to the broader
collaboration. Throughout all of these operations, we also desire to have a
robust automatic system that is not sensitive to poor quality data, and provides
monitoring tools for overall telescope operation as well as metrics about
individual array elements. The science goals of HERA---the detection and
characterization of the first stars and galaxies---are exquisitely sensitive to
sources of systematic errors, and so having real-time feedback on the
performance of the telescope is crucial. These separate but interrelated tasks
are accomplished primarily through the work of two packages: the
\texttt{hera\_opm} package\footnote{https://github.com/HERA-Team/hera\_opm},
responsible for running analysis on data, and the
\texttt{librarian}\footnote{https://github.com/HERA-Team/librarian}, responsible
for storing data and transferring to other sites. In the following discussion,
we provide a brief overview of the RTP system, and showcase some of its
features. In Sec.~\ref{sec:opm}, we discuss the processing performed by the
\texttt{hera\_opm} package. In Sec.~\ref{sec:librarian}, we describe the
\texttt{librarian}. Finally, in Sec.~\ref{sec:discussion}, we provide some
general discussion and mention planned future upgrades. All of the software and
tools used throughout this manuscript are free and open source software licensed
under the BSD-2-clause License, with the codebases freely available on
GitHub. The packages are designed to be flexible enough to accommodate the needs
of other telescopes, and more generally any data reduction pipelines.

\section{Processing}
\label{sec:opm}

A single visibility data file for HERA contains data corresponding to all
baselines, frequencies, and polarizations, for several consecutive times. In
general, we wish to perform the same fixed set of analysis routines on each data
file. Furthermore, there are several steps that are effectively prerequisite for
a subsequent step. For example, an initial calibration solution is determined
using a redundant calibration technique \cite{dillon_etal2020}, after which a
sky-referenced calibration is performed \cite{kern_etal2020}. The redundant
calibration must be performed before the absolute calibration, though only for
each datafile specifically. On the other hand, there are some analysis
techniques, such as smoothing of calibration solutions in time and frequency,
that require an earlier step to have completed for all files. Finally, there are
some steps that must operate on several files consecutive in time as a single
unit, such as identifying data contaminated by radio frequency interference
(RFI), and stride through the data files operating on chunks. From a user
perspective, the goal is to provide a list of tasks to run for each file, any
necessary prerequisites for the file (including ones adjacent in time), and
describe a method for chunking data up into steps. Once the workflow and input
data files are provided, the user should be able to dispatch the workflow to
appropriate compute resources automatically, without having to individually run
the steps for each file and keep track of dependencies by hand. Such an approach
helps automate the analysis process, and provides a reliable and reproducible
way to perform complicated workflows on a large volume of data.

The primary means by which these goals are accomplished in HERA is through using
the \texttt{hera\_opm} package, which provides the Online Processing Module
(OPM). There are several steps required for using this package. As mentioned
above, the user first defines a \textbf{configuration file} which contains the
workflow of steps to perform on each file. For each step, the user lists any
prerequisites, time chunking, options, and specific arguments to pass to shell
scripts. In addition, the user writes a series of \textbf{task scripts}, which
carry out the actual analysis steps. These are generally shell scripts, which
may in turn call other scripts or programs (Python or Perl scripts, compiled C
analysis programs, etc.). For several steps in HERA, these tasks execute
dynamically generated Jupyter Notebooks\footnote{https://jupyter.org/}, which
are automatically uploaded to GitHub for wider consumption. The tools in
\texttt{hera\_opm} convert the configuration file and a list of input files into
a series of wrapper scripts. In addition, a file is generated that is suitable
for input to the \texttt{makeflow} program \cite{makeflow}, part of the
\texttt{cctools}
package\footnote{https://github.com/cooperative-computing-lab/cctools}. The
\texttt{makeflow} file generated contains all of the rules necessary to execute
the workflow as defined by the user in the configuration file. Furthermore,
\texttt{makeflow} has the ability to interface with either local execution on a
user's personal machine, or deploy execution to a cluster resource manager. For
on-site processing in HERA, we opt for the latter, and make use of the
Slurm\footnote{https://slurm.schedmd.com/} scheduler installed on the local
processing cluster. However, \texttt{makeflow} allows for straightforward use of
different schedulers, which increases portability. For example, HERA performs
further post-processing analysis at National Radio Astronomy Observatory (NRAO)
facilities which uses
TORQUE\footnote{https://adaptivecomputing.com/cherry-services/torque-resource-manager/},
which is accomplished by changing a single line in the configuration file.

As part of on-site operations, a monitoring script waits for new data to be
generated. Once data collection is complete, a matching \texttt{makeflow} input
file and accompanying wrapper scripts are generated, which enables launching a
new workflow. The workflow is automatically launched, and runs by dispatching
jobs to the Slurm cluster on-site. As part of the workflow, data products such
as calibration solutions and flags corresponding to poor quality data are
generated. Some of these products, as well as some of the raw data, are saved as
part of our on-site cluster storage management system, and transported to remote
processing clusters. These actions are handled by the \texttt{librarian}
package, which we turn to in Sec.~\ref{sec:librarian}.

\section{Data Storage}
\label{sec:librarian}

In addition to processing data on-site, we also want to be able to store raw
data, calibration solutions, data quality metrics, and derived data products
indefinitely. Although simple indexing and cataloging schemes may be able to
keep pace for several thousand files and straightforward organizational
structures, the problem quickly becomes intractable when applied to millions of
files across hundreds of nights of observation, each with unique qualities that
may affect the overall results in different ways. For HERA, the 60 TB of data
each night will be divided into roughly 50,000 files, each of which will have
unique products and metadata associated with them. To accommodate such a
tremendous proliferation of files, we have developed the \texttt{librarian}
package, which handles tracking of data files and transporting them to remote
clusters. The underlying architecture also scales well with total number of
files to track, so that the data can be added and accessed without significant
slowdown as the number of files becomes large. We now briefly describe some of
the main features of this package, and describe how its functionality enables
the RTP system to provide data and data products to users working at remote
clusters.

The \texttt{librarian} consists of two components: a \textbf{librarian server},
which catalogs and stores data files, and the \textbf{librarian client}, which
supports access in various ways. The server runs a tornado-based web
server\footnote{https://github.com/tornadoweb/tornado}, and makes use of a
backing PostgeSQL\footnote{https://www.postgresql.org/} relational database. The
server also requires access to large storage drives (typically RAID arrays) for
saving data, called \textbf{stores}. However, these storage drives need not be
on the same machine as the \texttt{librarian} server. Additionally, the
\texttt{librarian} can make use of multiple stores across different machines. As
part of the internal bookkeeping, the \texttt{librarian} keeps track of which
store a file is saved to, so that it is able to access it again in the
future. The \texttt{librarian} client supports web-based access to the tornado
server as well as a command line interface for programmatic use of
\texttt{librarian} utilities. There is also a Python-based API for incorporating
\texttt{librarian} access into analysis scripts. The client access, either
web-based or command line-based, provides search capabilities for files known to
the \texttt{librarian} server, as well as the ability to copy data to a local
storage location or remote machine.

When a file is added to the \texttt{librarian}, the server reads some of the
metadata to determine which observation session the file corresponds to. The
server also performs file-level checking, such as computing the file's checksum
and inferring the file type. These bits of metadata are saved along with the
actual data, and copied to a store. The \texttt{librarian} automatically groups
multiple observations that are contiguous in time into a single observing
session, which is treated as a ``night'' of data. When searching for data, users
can specify the desired file name itself (or include wild-card characters to
match multiple files), the file type, individual observation identifier, or
observation session identifier. The \texttt{librarian} also supports searching
for combinations of these identifiers, allowing for straightforward access based
on a variety of features.

The \texttt{librarian} also supports communication with other \texttt{librarian}
servers installed at other locations. For example, \texttt{librarian} servers
run both on-site and at the NRAO processing facility that HERA uses. These
\texttt{librarian} servers are loosely coupled, and are able to transfer
metadata of individual files between each other. The general problem of syncing
files is rendered tractable by requiring that files have globally unique names,
and that both these names as well as the file contents are immutable. In
practice, the unique name constraint is easy to adhere to by using the Julian
date of observation as part of the file name. The inter-\texttt{librarian}
communication also supports automatic transfer of files meeting certain criteria
between servers. For example, all new data can be transferred from the on-site
\texttt{librarian} to the NRAO server, or just files corresponding to
calibration solutions, or only site-specific maintenance data, etc. The transfer
of files between \texttt{librarian} servers supports both simple transferring
using \texttt{rsync}, as well as more sophisticated transfer protocols such as
Globus\footnote{https://www.globus.org/}.

As an additional feature, the \texttt{librarian} installation at NRAO features
what is known as ``staging'' data from the long-term storage nodes to short-term
scratch space for user interaction. The user performs a search for specific
files, after which the user requests that the data be transferred to their
personal scratch space. The \texttt{librarian} automatically fetches the data
from the appropriate NRAO store and copies it to the user's scratch space, after
which the \texttt{librarian} notifies the user that the data was successfully
transferred (or of any errors that may have occurred). Accordingly, the user
need not worry about manually copying data, or by implementation details such as
files being saved across different local stores. In many cases, these high-level
access patterns facilitate user access to data without requiring a cumbersome
and error-prone series of command line calls, allowing users to focus on
analyzing data rather than hampering their access to it.

\section{Discussion}
\label{sec:discussion}

Successful operation of the RTP system on-site requires the coherent and
interconnected operation of both the processing and data storage services. There
are also a series of monitoring and control utilities that help ensure continued
operation of all of these services, though we defer detailed discussion of these
processes to future work. In this section, we briefly touch on the way in which
the processing and data management services interact with the raw data on-site,
and explore future planned improvements. We also highlight ways in which these
tools may be adapted to operation of other telescopes or data reduction
pipelines.

\begin{figure}
  \centering
  \includegraphics[width=0.49\textwidth]{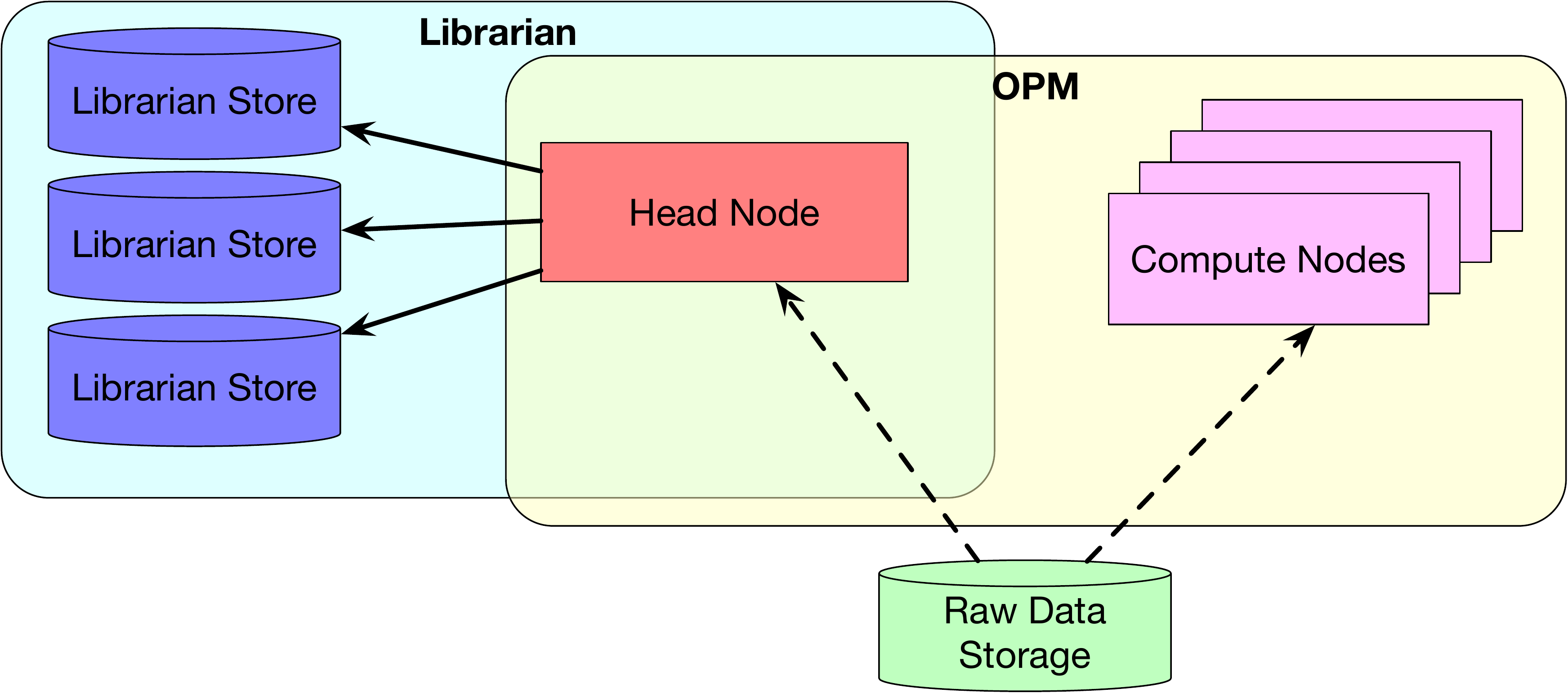}
  \caption{A schematic of how the RTP system operates on-site as part of the
    HERA processing cluster. Both the \texttt{librarian} and \texttt{hera\_opm}
    systems make use of the head node, as well as auxiliary machines. See
    Sec.~\ref{sec:discussion} for additional discussion of how the systems
    interact.}
  \label{fig:rtp}
\end{figure}

Figure~\ref{fig:rtp} shows a schematic of how the different processes interact
on-site. The main head node of the cluster runs the \texttt{librarian} server
process, as well as the monitoring service waiting for acquisition of new data,
represented in the diagram as the ``Raw Data Storage'' location. The storage
disk on this machine is mounted via NFS to the head node as well as all of the
compute nodes, which allows for simple access without needing to explicitly copy
data to local storage. As part of the workflow which processes the raw data,
several steps include uploading files to the \texttt{librarian} server running
on-site. Currently the data volumes for HERA are sufficiently modest that all
raw data and derived products can be uploaded and transferred, though as the
number of active antennas rapidly expands in the coming year, this will no
longer be possible. When new data are uploaded to the \texttt{librarian}, they
are automatically transferred to the store with the most free space available,
to avoid overfilling a single store. Transfer of uploaded data files to the NRAO
\texttt{librarian} is handled automatically, and proceeds as data are
available. At present, the system is able to keep up with processing
requirements in real time, though as the data volume grows additional testing
and benchmarking will be required to ensure that the RTP remains real time, and
can keep pace with the larger computational load required.

In the future, we plan to make use of other computational resources available to
HERA, beyond those which are located on-site. In South Africa, the
Inter-university Institute for Data Intensive Astronomy (IDIA) hosts
computational resources that are available to high-performance astronomy
applications. HERA has computational resources available as part of the IDIA
community, and plans to run some portion of its processing load on IDIA
machines. Future development will be required to leverage these resources
efficiently, and fold the different steps at different resources into a single,
coherent workflow. Nevertheless, given the dramatic increase in data volume and
associated computational requirements in the near future, HERA will need to make
use of all available resources.

Finally, we would like to emphasize that these processing tools are freely
available, and can be adapted for use as part of other astronomy applications
and workflows. Although the \texttt{hera\_opm} and \texttt{librarian} packages
were initially developed to work in the context of HERA processing and data
management, they are written in a sufficiently general manner to work in other
contexts. For example, the \texttt{librarian} is currently being adapted for use
in the Simons Observatory\footnote{https://simonsobservatory.org/}, an
experiment observing the cosmic microwave background. Similarly, the core
functionality of the \texttt{hera\_opm} package is not specialized to HERA data,
and can be used generally for any software pipeline applied to a series of input
files. The software powering these packages is free and open source, is
available on GitHub, and has an active team of developers behind them. It is our
hope that other groups find them useful, and can use them for their own data
processing needs. For questions and comments about the use of these systems,
please reach out to us via email or GitHub.

%
% Note that the authors' affiliations and contact information should be
% here, after the references.
%

\noindent\small
Paul La Plante is with University of California, Berkeley,
CA 94720, USA; e-mail: plaplant@berkeley.edu.\\[1ex]
Peter K. G. Williams is with Harvard-Smithsonian Center for Astrophysics, Cambridge,
MA 02138, USA,
and the American Astronomical Society, Washington,
DC 20006, USA; e-mail: pwilliams@cfa.harvard.edu.\\[1ex]
Joshua S. Dillon is with University of California, Berkeley,
CA 94720, USA; e-mail: jsdillon@berkeley.edu.


\begin{thebibliography}{9}

\bibitem{deboer_etal2017} D.~R.~DeBoer, A.~R.~Parsons, J.~E.~Aguirre,
  P.~Alexander, Z.~Ali et al., ``Hydrogen Epoch of Reionization Array (HERA),''
  \emph{Publications of the Astronomical Society of the Pacific}, \textbf{129},
  974, March 2017, 27 pp.

\bibitem{wijnholds_etal2018} S.~J.~Wijnholds, A.~G.~Willis, and S.~Salvini,
  ``Baseline-dependent averaging in radio interferometry,'' \emph{Monthly
    Notices of the Royal Astronomical Society}, \textbf{476}, May 2018,
  pp. 2029-2039.

\bibitem{dillon_etal2020} J.~S.~Dillon, M.~Lee, Z.~S.~Ali, A.~R.~Parsons,
  N.~Orosz et al., ``Redundant-Baseline Calibration of the Hydrogen Epoch of
  Reionization Array,'' \emph{arXiv:2003.08399}, March 2020, 23 pp.

\bibitem{kern_etal2020} N.~S.~Kern, J.~S.~Dillon, A.~R.~Parsons, C.~L.~Carilli,
  G.~Bernardi et al., ``Absolute Calibration Strategies for the Hydrogen Epoch
  of Reionization Array and Their Impact on the 21\,cm Power Spectrum,''
  \emph{The Astrophysical Journal}, \textbf{890}, 2, February 2020, 24 pp.

\bibitem{makeflow} M.~Albrecht, P.~Donnelly, P.~Bui, and D.~Thain, ``Makeflow: A
  Portable Abstraction for Data Intensive Computing on Clusters, Clouds, and
  Grids,'' \emph{SWEET '12: Proceedings of the 1st ACM SIGMOD Workshop on
    Scalable Workflow Execution Engines and Technologies}, New York, NY, May
  2012, pp. 1-13.

\end{thebibliography}
\end{document}